\documentclass[2p]{elsarticle}
\usepackage{natbib,longtable}
\usepackage{graphicx,array}

\usepackage{appendix}   

\begin{document}

\begin{frontmatter}

  \title{Evaluated Experimental Isobaric Analogue States from $T = 1/2$  to $T = $ 3 and associated IMME coefficients}

  \author[IPNO]{M. MacCormick\corref{cor1}}
  \ead{E-mail: maccorm@ipno.in2p3.fr}

  \author[CSNSM]{G. Audi}

  \cortext[cor1]{Corresponding author.}
 
  \address[IPNO]{Institut de Physique Nucl\'eaire CNRS/IN2P3 and Universit\'e Paris-Sud,Orsay, France}

  \address[CSNSM]{Centre de Sciences Nucl\'eaires et de Sciences de la Mati\`ere CSNSM, CNRS/IN2P3 and Universit\'e Paris-Sud,Orsay, France }

\begin{abstract}
Isobaric multiplets can be used to provide reliable mass predictions through the Isobaric Multiplet Mass Equation (IMME). Isobaric analogue states (IAS) for isospin multiplets from $T$~=~$1/2$ to $3$ have been studied within the 2012 Atomic Mass Evaluation ({\sc Ame2012}). Each IAS established from published experimental reaction data has been expressed in the form of a primary reaction $Q$-value and, when necessary, has been recalibrated. The evaluated IAS masses are provided here along with the associated IMME coefficients. Quadratic and higher order forms of the IMME have been considered, and global trends have been extracted. Particular nuclides requiring experimental investigation have been identified and discussed. This dataset is the most precise and extensive set of evaluated IAS to date. 
\end{abstract}


\end{frontmatter}

\section{Introduction}
Discovering and understanding basic symmetries in nature is of fundamental importance in modern physics. This work focuses on isospin symmetry and the study of isospin-dependent nuclear masses via isobaric analogue states (IAS). Exploring the simple relationship between these states allows to extract information on nuclear Coulomb effects, the charge independence of nuclear forces, and nuclear masses. The abundance of high precision experimental data available, accumulated over an extensive set of nuclides, are an essential source of experimental input to several areas of nuclear physics. 

Modelling stellar evolution requires the knowledge of low rate weak interaction rates~\cite{2006Zegers}, which can be examined via accelerator based charge-exchange reactions. The comparison between Gamow-Teller decays and the mirror IAS Fermi decays allow to extract the GT strength transition. The equation of state of dense symmetrical nuclear matter, another essential component of realistic astrophysical modelling~\cite{20012Tsang} uses IAS experimental data to constrain the symmetry energy component~\cite{2013Danielewcz}. Historically, J.C.~Hardy and I.S.~Towner~\cite{2009Hardy12} have used IAS masses to test the Conserved Vector Current hypothesis of the vector part of the weak interaction, an open, as yet unresolved quest spanning more than 50 years. Finally, IAS are used in the modelling of nuclear mass where analysis of these states provide valuable, accurate estimates of unmeasured masses.

In this paper we provide the most recent and evaluated set of Isobaric Analogue State mass excesses based on the evaluation in {\sc Ame2012}~\cite{AME2012a, AME2012b}, {\sc Ensdf}~\cite{ENSDF} and {\sc Nubase2012}~\cite{NUBASE2012}. Values of the quadratic, and when possible, higher order, Isobaric Multiplet Mass Equation (IMME) coefficients are then extracted. After considering the robustness of the IMME predictions the quadratic coefficients are used to predict currently unmeasured IAS excited state mass excesses. Finally, the best fits to data are considered in the framework of first order perturbation theory and the global $b$ and $c$ tendencies over the most extensive range of atomic masses are shown. 

The study of Isobaric Analogue States (IAS) and the Isobaric Multiplet Mass Equation (IMME) have been the subject of continuous effort since the late '50's, but is less in the foreground as it used to be. It is therefore worth briefly recalling the basis on which these studies are carried out.

\section{Isospin}
Heisenberg introduced the new isospin quantum number $T$ via the concept of the nucleon which is defined to have two states~\cite{Heisenberg}, the neutron with $T_{z}=+1/2$ and the proton as $T_{z}=-1/2$ (using today's experimental convention). The isospin quantum number follows the same mathematical rules as for spin, where each nucleon has a projection which is either aligned or anti-aligned with the z-axis. For any nucleus of $A$ nucleons, the good quantum number is the isospin projection $T_{z}$ as defined by the quantity 
\begin{equation}
 T_{z} = \frac{N-Z}{2}
\label{eqnTz}
\end{equation}
\noindent
where $N$ is the number of neutrons and $Z$ the number of protons. Each nuclide can then have nuclear configurations where the isospin $T$ is limited to the set of possibilities
\begin{equation}
 |\frac{N-Z}{2}| \leq T \leq \frac{N+Z}{2}
\label{eqnT}
\end{equation}
\noindent Consequently, isobaric nuclides which have the same isospin value $T$ form an isobaric multiplet~\cite{IASbasics}. 

\subsection{Nuclear mass}
In the charge independent description, nuclear mass can be described as being the sum of three components:
\begin{equation}
 M(A, T, T_z) = M_0 + E_C + \Delta_{nH}T_z 
\label{eqnMass}
\end{equation}
where $M_0$ is some basic nuclear mass, $E_C$ the Coulomb energy and $\Delta_{nH}$ the neutron-$^1$H mass difference.
Using first order perturbation theory, and considering the presence of a two-body charge-symmetry breaking force to describe the neutron-proton force, the Coulomb energy $E_c$ may be written as
\begin{equation}
 E_c(A, T, T_z) = E{_C^{(0)}}(A, T, T_z) + T_z E{_C^{(1)}}(A, T) + T{_z^2} E{^{(2)}_C}(A, T)
\label{eqnCoulomb}
\end{equation}
where $E{_C^{(0)}}$, $E{_C^{(1)}}$ and $E{^{(2)}_C}$ are isoscalar, isovector and isotensor parts of the Coulomb energy~\cite{CoulombEnergy}.

In this charge independent view of the nuclear force, the energy levels of the multiplet members, as calculated using the instrinsic spins and relative angular momenta, lead to the conclusion that the multiplet members must also have the same spin and parity attributions. 

\subsection{The IMME}
E.P.~Wigner suggested that the energy difference between members of an isobaric multiplet $T$ follows a quadratic form:
\begin{equation}
a + b T_{z} + cT^{2}_{z}
\label{eqnIMME}
\end{equation}
where the coefficients $a$, $b$ and $c$ are interpreted as being the scalar, vector and tensor parts of a charge independent nuclear Hamiltonian. 
Using the Coulomb energy description in equation~(\ref{eqnCoulomb}) the coefficients are identified to be:

\begin{equation}
a = M_0 + E{_C^{(0)}} - T(T+1)E{_C^{(2)}}~~~~~b = \Delta_{nH} - E{_C^{(1)}}~~~~~c = 3E{_C^{(2)}}
\label{eqnCoeffHam}
\end{equation}

The $a$ coefficient is simply the mass excess of the $T_z$=0 member for integer multiplets, and is rather difficult to interpret for the half-integer cases.  The $b$ coefficient is the dominant term in the IMME giving, to first order, the general slope of the mass multiplet parabola, leading to energy differences proportional to $T_{z}$. However, if higher order effects are non-negligible, and arise from the interaction between neighbouring isospin states differing by one isospin unit, it is this term that is expected to reflect the amount of mixing~\cite{IMMEpaper}. The $c$ coefficient describes the strength of the interaction between states separated by 2 isospin units and may also contain higher order terms if first order theory is insufficient.

\subsection{IMME limits}
Isospin is a useful symmetry based concept for mainly low lying, well defined levels in nuclei from $A\simeq10$ up to $A\simeq60$, and in high energy collective excited states of well defined isospin. However, it has its limits; in intermediate mass nuclei at higher excitation energies the effects of isospin mixing due to the growth in the Coulomb component cannot be ignored, and the basic first order perturbation theory behind the IMME is no longer valid, isospin consequently becomes a less useful quantity~\cite{Wigner91}. At higher mass, beyond $A=60$, the charge-independence hypothesis is no longer valid for such a large collection of protons in the nucleus. However, analogue resonant states in even heavier nuclei ($A>80$) and at higher excitation energies are again expected to show simple isospin symmetry since the higher excitation energy, implying shorter nuclear interaction times~\cite{IMMEbasics}, leads to fewer possibilities for isospin mixing, as observed in isobaric analogue resonances. A.M.~Lane and J.M.~Soper reviewed this subject in detail in the classical reference~\cite{LaneSoper} and  explain why isospin may be a good quantum number, but essentially ``useless" in heavy nuclei since its degree of purity cannot be directly checked by experiment. 

The IMME as proposed by Wigner supposes a simple nuclear model of constant density, equal radii between IAS states, and the same spin-parity for the multiplet members. Even though the nuclear description is greatly simplified, the IMME still provides a reasonable mass estimate for nuclei up to $A\sim60$. In general, when the estimated IMME value and the experimental results disagree, there is experimental evidence for a greater probability of isospin mixing, leading to the fragmentation~\cite{MacDonald} of the IAS over two, but occasionally more, states. 

\section{Data Selection} 
There are two types of data sources used in this evaluation; nuclear reaction data included in {\sc Ame2012}, and nuclear spectroscopy data, as evaluated in {\sc Ensdf}. Sets of isobaric multiplets constrained by two ground state nuclei are considered. A recent compilation of ground and excited IAS can be found in the appendix of reference~\cite{YiHuaLam} and is an update to the extensive work of M.S.~Antony and A.~Pape~\cite{IAS1984, IAS1985, IAS1986, IAS1998}.

The experimental reaction data published in refereed journals relates excited IAS levels to the ground state mass of a neighbouring nuclide  through the reaction $Q$-value, consequently they contribute to the evaluation of ground state masses, hence their inclusion in {\sc Ame2012}. The detailed procedures used in the AME will not be repeated here and can be found in the recent papers~\cite{AME2012a, AME2012b}. We recall that the reaction $Q$-value, for example, of the reaction A(a,b)B* where B* is an excited IAS state, provides a unique relationship between the masses in the initial $M_A$, $M_a$ and final states $M_b$, $M_{B^*}$, and is written:
\begin{equation}
Q = M_A + M_a - M_b - M_{B^*} 
\label{Qvalue}
\end{equation}

Experimental decay data evaluated by the {\sc Ensdf} group provides detailed information on internal nuclear transitions. This type of data relates excited states to the ground state of {\it the same} nuclide, for example, through B*($\gamma$)B. The experimental data comes dominantly from gamma spectroscopy measurements, with precisions that are usually at least two orders of magnitude better than that of reaction data. Almost all IAS threshold measurements are from (x,$\gamma$) experiments and so measure the decaying IAS state at production with excellent precision.

Since the aim here is to construct the most precise evaluated IAS dataset, decay data is used whenever available. However, in $107$~cases, data for the excited IAS is only available through nuclear reactions, and occasionally the selection rules are such that the IAS can only be observed experimentally via particle decays. Currently there are $13$~cases which are identified exclusively via reaction mechanisms and for which no equivalent is found in {\sc Ensdf}. Each case is labelled accordingly in the IAS tables in~\ref{IAStables}. 

\subsection{Rebuilding $Q$-values}
In many experiments, even though the reaction $Q$-value is the primary physical quantity measured, it is not always given explicitly in the final publication. This is more often the case when the scientific objective is to {\it deduce} excitation energies, and as such, knowledge of ground state masses is required. However, these masses are susceptible to small variations over the years as the mass evaluations evolve. Consequently, in cases where the authors do not directly give the experimental $Q$-value, a detailed search for the known mass excesses before, but closest to the time of publication, was carried out. In the absence of explicit values, authors generally cite the source of the mass excesses used to construct the final published results. When no source is provided, the {\sc Ame} nearest but prior to the publication date is used. 

In other cases authors provide the ground state reaction $Q$-value used at the time of the experiment and the IAS $Q$-value is reconstructed using the author's original reference data.  In the current evaluation {\it all} non-ground state IAS were rebuilt and compared to the author's original results.

\subsection{Recalibrations}

A few older experiments report $Q$-values that were calibrated with respect to ``well known standards" which have since shifted by small amounts, inducing an offset in the resultant $Q$-values. The main dataset subject to these displacements was published by Becchetti~{\it et. al.} in 1971~\cite{1971Becchetti29} and many unique IAS results were obtained for stable isotopes ranging from $A=42$ to $A=70$. Close inspection of the four calibration channels used in the original work show shifts varying from $-1$~keV to $+8$~keV using recent values. This dataset was recalibrated using the recipe that follows. 

The aim of the original experiment was to measure the Coulomb Displacement Energy (CDE)~\cite{CDE} for stable isotopes from $A=42$ to $A=70$, with a precision of $\pm6$~keV. $(^3He,t)$ reaction $Q$-values were measured on the focal plane of an Enge split-pole, the reaction being carried out on several targets. The focal plane was calibrated using the threshold value of $(p,n)$ reactions~\cite{1965Freeman08} on the same set of targets, and most measurements were carried out using the same detector position and magnetic field setting. The relationship between $(^3He,t)$ and $(p,n)$ ground state reactions is
\begin{equation}
Q({^3}He,t)_{meas} - Q(p,n)_{ref} = 764~keV
\end{equation}
where $Q(p,n)_{ref}$ was given in the original 1971 paper.  
The IAS reaction $Q$-values are then
\begin{equation}
Q_{IAS} = 764 - \Delta E_c
\end{equation}
where $\Delta E_c$ is the Coulomb displacement energy.

The $(p,n)$ threshold values have since evolved, as have the mass excesses. For example, consider the reaction $^{42}$Ca(p,n)$^{42}$Sc where, following~\cite{1965Freeman08} the threshold proton energy $E_p$ was measured at $7387.5~\pm~2.3$~keV. Using the non-relativistic approximation and recent mass values, correcting for recoil gives a $Q$-value of $-7214.24~\pm~2.25$~keV exactly the same as in 1965. However, the best currently available $(p,n)$ $Q$-value is calculated directly from the recent mass evaluation, at $-7208.446~\pm~0.097$~keV, a value more than one order of magnitude more precise than the 1965 experimental values, and offset by some $+5.792$~keV. This extra shift is applied to the dataset calibrated on the $\rm^{42}Ca(^3He,t)^{42}Sc$ reaction. Given that the experimental precision estimated at the time was $\pm6$~keV, this constitutes a non-negligible adjustment.  This procedure was repeated for the other calibration references used in the original publication and have associated comments in the final tables~\cite{AME2012a}.  

\section{IMME exception}
\subsection{$A=16, T=1$}

This unusual situation was first reported in 1956~\cite{1956Wilkinson} and is the only multiplet not delimited by two ground state masses. The triplet formed by $^{16}$N, $^{16}$O$^i$, $^{16}$F does not fit the usual isospin multiplet criteria due to asymmetric spin-parity characteristics of the ground state ``mirror" nuclei $^{16}$N and $^{16}$F. $^{16}$N is 2$^-$ and $^{16}$F is 0$^-$ and so cannot be used to build an isobaric multiplet where the same J$^{\pi}$ is a requirement for all members. This unique situation can be resolved by considering the $0^-$ isomeric state in $^{16}$N, $^{16}$N$^{m}$ at $120$~keV above ground state. The multiplet is exceptionally composed of the $0^-$ $^{16}$N$^{m}$, $^{16}$O$^i$, $^{16}$F triplet. The influence on the IMME fitted results is to lower the isovector component of the Couloumb energy by $60$~keV and to raise the isotensor component by the same amount. 

\section{IMME fits and results}
A few simple data selection rules, discussed in the next paragraph, were applied to the datasets before fitting the muliplets and extracting the IMME coefficients. Quadratic, and higher order IMME coefficients are tabulated in~\ref{IMMEtables}. 

\subsection{Data selection}\label{dataSelection}

Establishing the degree of isospin purity of nuclear levels requires detailed measurements of all formation and decay channels. Energy levels of the same spin-parity but, most probably, neighbours in isospin may interfere constructively and/or destructively, resulting in the displacement of the hypothetical pure levels to fragmented, isospin-mixed, states. Fragmentation is observed more readily for $A>40$ and at higher excitation energies, where the greater density of levels increases the probability of inducing this effect. Fragmented levels are also seen to occur in very lightweight nuclei, where shell effects play a more dominant role in each of the individual multiplet members. Higher order terms in both the $b$ and $c$ coefficients are thought to reflect the existence of internally fragmented states~\cite{IMMEpaper}. Experimentally observed fragmented levels have been measured in the $T=2$ multiplets for $A=44, 48, 52, 56$ and $60$. In the $A=56$ multiplet in particular, all the excited states are fragmented. Consequently, it is not possible to simply apply the IMME since there is an inherent one-mass to one-IAS supposition. The unfragmented masses may be used to predict the position of a hypothetical unfragmented pure isospin state and its position with respect to the experimentally observed fragmented energy levels may be useful in guiding theoretical calculations. Isospin mixing may also be the consequence of higher order Coulomb effects, showing up experimentally via higher IMME terms, and the study of Coulomb displacement energies extracted from neighbouring IAS is a valuable source of experimental information for theoretical studies~\cite{1983Auerbach}.     

In the particular case of the $T=1, A=8$ triplet, which includes the very exotic $^8$Be$^i$ $2^+$ excited state, spectroscopic studies~\cite{2004Tilley06} show that this state is fragmented into two, $T=0$ and $T=1$ isospin-mixed components separated by $296\pm3$~keV making the IMME difficult to apply. The treatment of fragmented states constitutes an ongoing study and requires a mixture of experimental and theoretical input to go further.

In the present study all known fragmented states were excluded from the IAS fitting procedures, with the exception of $^{52}$Fe$^j$  where there may be a fragmented doublet separated by 4 keV, which is smaller than the mass precision of $\pm$6~keV.

\subsection{Data fitting}
Even if higher order terms provide a better fit to the data, the first three $a$, $b$, $c$ coefficients should remain relatively stable. That is, the higher order corrective terms should be added to the basic function, and so the mass $M$:
\begin{equation}
M = a + bT{_z} + cT{_z}^2 
\end{equation}
and
\begin{equation}
M = a + bT{_z} + cT{_z}^2 + dT{_z}^3 
\end{equation}
should have equivalent $a$, $b$ and $c$ values within the general tendencies, then, the $d$ coefficient can be considered to be the amplitude of a higher order correcting term. When possible, tests for non-zero $eT{_z}^4$ terms were also tried. A non-zero coefficient is defined with respect to a $\pm$3$\sigma$ limit, for example $d = 0.15\pm 0.06$ is compatible with zero, whereas $0.15\pm0.01$ implies a non-zero value. All results are shown in the final tables in~\ref{IMMEtables} but the interpretation of results in the text uses this basic criterion. 

\subsection{Precision and Convergence}\label{PrecAndConv}
An unusual situation occurs in the $T=3/2$ series for $A=41$ where the very high data precision on $^{41}$K, of $\pm0.004$~keV,  ``fixes" this data point, effectively reducing the fit by one degree of freedom. To obtain a fully converged fit for the coefficients and, more importantly in this case, the associated errors, the $^{41}$K experimental error bar was empirically changed to $0.01$. The central fit values remain unchanged by this adjustment and the IMME coefficient errors are considered to be artificially increased. A higher order fit requires even more freedom in the experimental precision to converge, so, although $a$, $b$, $c$ and $d$ coefficients could be extracted, they do not reflect the high precision of the data. It is concluded that the higher order model is not a good fit to the data.  

Another unusual case is the $T=2, A=40$ quintuplet where there is simultaneously one very precise ground state value for $^{40}$Ar at 
$\pm0.0022$~keV and a relatively less well known mass excess for $^{40}$Ti with an error bar at $\pm160$~keV. Changing the error on $^{40}$Ar allows the error bars of the second order IMME fit to converge, but is still too stringent for higher order approaches to succeed. When the $^{40}$Ar error is artificially increased to $\pm0.05$~keV, the higher order fits converge, but the $d$ coefficient then works out at $0.34\pm1.17$~keV, and is negligible within a $\pm3\sigma$ criteria. The $T=2, A=40$ multiplet is therefore best described by a second order IMME and the associated fit errors are considered to be artificially increased. A more precise measurement of the $^{40}$Ti ground state mass would be of use in this particular case. 

\subsection{Individual cases}
\subsubsection{$T=1, A=44$}\label{Teq1Aeq44}

The fit to the $^{44}$Sc, $^{44}$Ti$^i$, $^{44}$V (J$^{\pi}$=2$^+$) triplet returns a doubtful negative $c$ coefficient of $-26\pm91$~keV. 
The original $^{44}$V mass excess data of $-23980+80-380$~keV~\cite{2004Stadlmann05} has been evaluated in {\sc Ame2012} at a value of $-24120\pm180$~keV, with symmetrized error bars. The experimental resolving power was insufficient to separate the $^{44}$V ground state from a suspected isomeric state, inducing a systematic error of $270\pm100$~keV. It is highly probable that this is the source of the unusual negative valued $c$ coefficient. Estimations are made for this case in section~\ref{predict}. 

The  possibility of fragmentation of the $^{44}$Ti$^i$ state is unlikely, with no observed fragmented $T=1$ states at higher masses. This can be understood from the generally lower excited states involved in these multiplets. 

The current case remains unsolved and a more precise measurement of the $^{44}$V ground state is necessary to move forward. 

\subsubsection{$T=3/2, A=9$}\label{Aeq9}
The $A=9, 3/2^-$ multiplet has an unambiguously non-zero $d$ coefficient and has been discussed in detail recently~\cite{2012Brodeur}. A shift in the energy levels of $^9$B$^i$ and $^9$Be$^i$ due to isospin mixing with a nearby $T=1/2$ state is shown to reproduce the experimental observations. A detailed spectroscopic study of $^9$B and $^9$Be focussed on the identification of $T=1/2$ and $T=3/2$ isospin mixed fragments, and the relative strengths, would be of great interest in resolving this case.   

\subsubsection{$T=3/2, A=11$}\label{Aeq11}
This quadruplet is listed in table~\ref{tab:J2} as having four experimentally defined masses. The quadratic function returns results out-with the normal smooth systematic tendencies. If all four data points are fitted, the $T_z=-1/2$ component is calculated to be $170$~keV below the general curve, more than $4\sigma$ away from the expected value. The inclusion of a cubic term significantly alters the $b$ and $c$ coefficients, by $-200$~keV and $-60$~keV, respectively. This casts doubts on the $^{11}$C$^i$ and $^{11}$B$^i$ masses. 

Eliminating $^{11}$C$^i$ from the fit results in a $c$ value of 162$\pm$10~keV, which is compared to 58$\pm$26~keV obtained by removing $^{11}$B$^i$ from the fit. Using the latter result implies an estimated mass excess for $^{11}$C$^i$ of 22602~keV, more than 200~keV below observations. Consequently, it is concluded there is an unresolved problem within this mutliplet and the problem most probably resides in the identification of the T$_z$=$-1/2$ $^{11}$C$^i$ IAS. The current $^{11}$B$^i$ IAS mass is retained since it results in IMME coefficients within the general tendencies.  This question was explored in~\cite{2001SherrFortune} where it was concluded that ``the ${\frac{1}{2}}^+$, $T=\frac{3}{2}$ states in $^{11}$B and $^{11}$C have been misidentified". A more recent analysis~\cite{2012Fortune} points to similar conclusions as here, ``that the excitation energy in $^{11}$C should be about 200~keV" lower.  
  
New experimental data for $^{11}$B$^i$, where the attributed spin-parity is still ambiguous, and a more precise determination of $^{11}$C$^i$, would be of benefit, the priority being on $^{11}$C$^i$.

\subsubsection{$T=3/2, A=53$}\label{Aeq53}
Although the $A=53$ multiplet returns a non-zero $d$ coefficient, the precision of the other coefficients is degraded and the central values shifted significantly when compared to the second order approach. Since the third order fit does not result in a simple correction to the lower order case, the second order fit is taken to be the best description of the data. 

\subsubsection{$T=3/2, A=55, 57, 59$}
Several states are observed to be fragmented and the remaining unfragmented data for each multiplet are not sufficient for second order fits.

\subsubsection{$T=2$}
The $T=2$ quintuplet measurements range from $A=8$ to $A=60$, and after the data selection (discussed in section~\ref{dataSelection}), there remains 16 datasets, of which 13 are 4-nucleon $(4n)$ systems and three are $(4n+2)$. Up to $A=40$ most of the $(4n)$ multiplets are complete, with the exception of $A=16$ where only four points are known experimentally. The missing data is experimentally difficult to access, corresponding to an extremely short-lived particle-unbound state.

\subsubsection{$T=2, A=8$}\label{Aeq8}
The $A=8$ multiplet has the unique distinction of having significant $d$ and $e$ coefficients, so including a fourth order correcting term. The cubic $d$ coefficient is 9$.42\pm1.88$~keV with a reduced $\chi^2=0.600$ and a $44\%$ probability of being the correct model for the data.  Allowing for a higher order term, leads to $d= 8.47\pm0.14$~keV and $e= 1.40\pm0.07$~keV. The precision of the fit for the other coefficients is an order of magnitude higher, with the $bT_z$ and $cT{{_z}^2}$ terms conceding $2$~keV and $4$~keV to accommodate the extra $eT^4$ term. So, although the fourth order fit is fully constrained, it also appears to provide the best fit to the data. 

This system has been the subject of recent discussion~\cite{2011Charity53} where a different $^8$C ground state mass is used. A theoretical and experimental investigation of this unusual case would be of help in identifying any potential problems with what appears to be an exceptional result. Higher precision measurements of $^8$C and $^8$B would also clarify whether or not this result is merely an artefact of the experimental resolution. 

\subsubsection{$T=2, A=32$}\label{Aeq32}
The $A=32$ result stands apart as being the unique example where a third order fit has a 96\% probability of being the correct model for the data. The $d$ coefficient amounts to some $0.89\pm0.11$~keV, an order of magnitude smaller than that of the $A=8$ multiplet. 

All of the excited states are evaluated by {\sc Ensdf} and, since 2006, the masses have significantly shifted by $-3.17, -0.35, -0.17, -3.1, 0.2$~keV going from $T_z=2$ to $T_z=-2$ ($^{32}$Si to $^{32}$Ar). The second order fit fails with a $\chi{^2}/n=31.7$ and less than $0.1\%$ probability of being the correct model. The fourth order fits give non-zero $d$ and $e$ coefficients, but do not provide a better fit to the data. They are statistically insignificant considering the associated errors, and are eliminated based on the $\pm3\sigma$ selection criterion.  

The source of non-zero $d$ and $e$ coefficients is not unambiguously resolved, and recent theoretical discussion~\cite{2011Signoracci} points again to isospin mixing as being a more probable generator of higher order coefficients than higher order Coulomb and charge effects alone~\cite{2012Brodeur}. 
\begin{figure}[t]
\centering 
\includegraphics[height=9cm,clip]{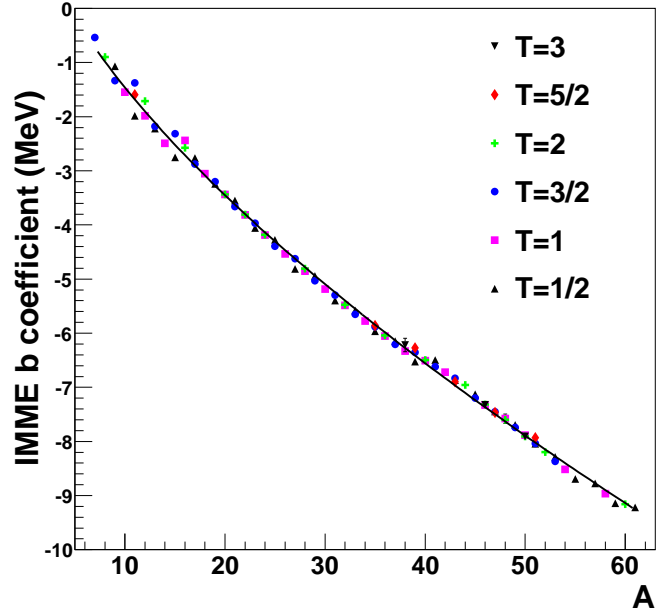}
\caption{IMME $b$ coefficients extracted from best fits. The full line is the global averaged function $b = \Delta_{nH} - 0.979 K + 0.702$~MeV where $K=0.720\times (A-1)/A{^{1/3}}$~MeV, inspired directly from the homogeneously charged spherical model.\label{fig:bCoeffsAverageLineNoE} }
\end{figure} 
\subsubsection{$T=2, A=52$}\label{Aeq52}
The four experimentally determined states includes $^{52}$Co$^i$, identified through its proton decay to $^{51}$Fe~\cite{2007Dossat17}. In the original work, the authors doubt the value attributed to the $^{52}$Co ground state when deducing the IAS excitation energy, but do not exclude the possibility of missing $\gamma$ decay energy in the experimental measurement. In this work, when this data point is included in the IMME fit, the quadratic $c$ coefficient result is unusual, at $c=156.36\pm3.31$~keV, roughly $25$~keV below the general tendencies. When a third order term is included the $c$ coefficient falls to $101\pm10$~keV and returns a non-zero $d$ coefficient of $28.82\pm4.47$~keV. The higher order term is not considered as a valid correction to the IMME since the $c$ coefficient is not stable. However, using only the other three data points, which includes the pivotal $T_z$=0  $^{52}$Fe$^j$ state, returns coefficients in keeping with the overall trends. $^{52}$Co$^i$ is excluded from the final IMME fit, however it is included in the IAS table~\ref{tab:J3}. The predicted $^{52}$Co$^i$ mass is discussed in section~\ref{predict}.

\begin{figure}[h] 
\centering
\includegraphics[height=9cm,clip]{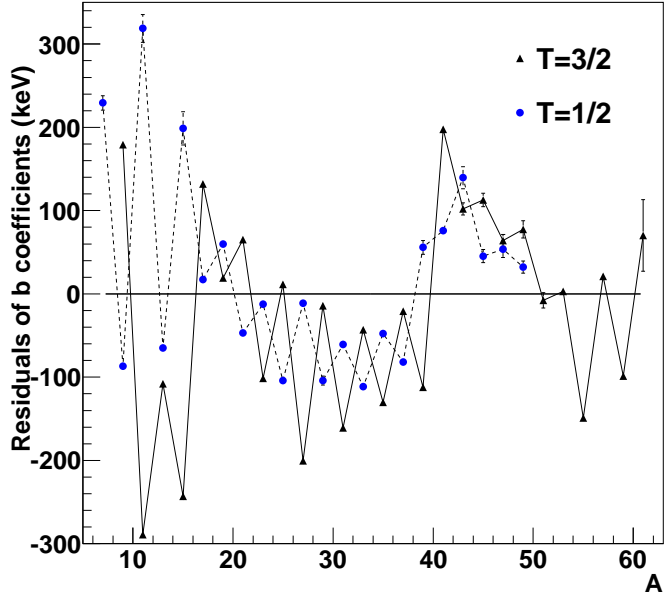}
\caption{Pronounced ``staggering"  effect that naturally appears in the IMME $b$ residuals. The residual values shown are calculated with respect to an averaged linear function: $b=0.720\times A^{2/3}+1.868$~MeV. \label{fig:unseparatedbStaggeringNoE} }
\end{figure}
\subsubsection{$T=3, A=54$\label{Aeq54}}
A particular case that was rejected from the IAS fitting procedure is the $T=3, A=54$ multiplet consisting of three neutron-rich experimentally determined members $^{54}$Cr, $^{54}$Mn$^i$ and $^{54}$Fe$^j$, and no observed fragmented states. However, the most excited T$_z$=0 state in $^{54}$Co defines the IMME $a$ coefficient and its absence, coupled with the lack of information on the proton-rich side implies that the IMME shape is not sufficiently well defined for a convincing IMME fit.

A measurement of the highly excited $T=3$ $^{54}$Co$^k$ state and of $^{54}$Ni$^j$, $^{54}$Cu$^i$ or the very exotic $^{54}$Zn would be a unique contribution to the currently sparse data.

\section{Global IMME coefficients}\label{globalIMME}
\begin{figure}[h] 
\centering
\includegraphics[height=11cm,clip]{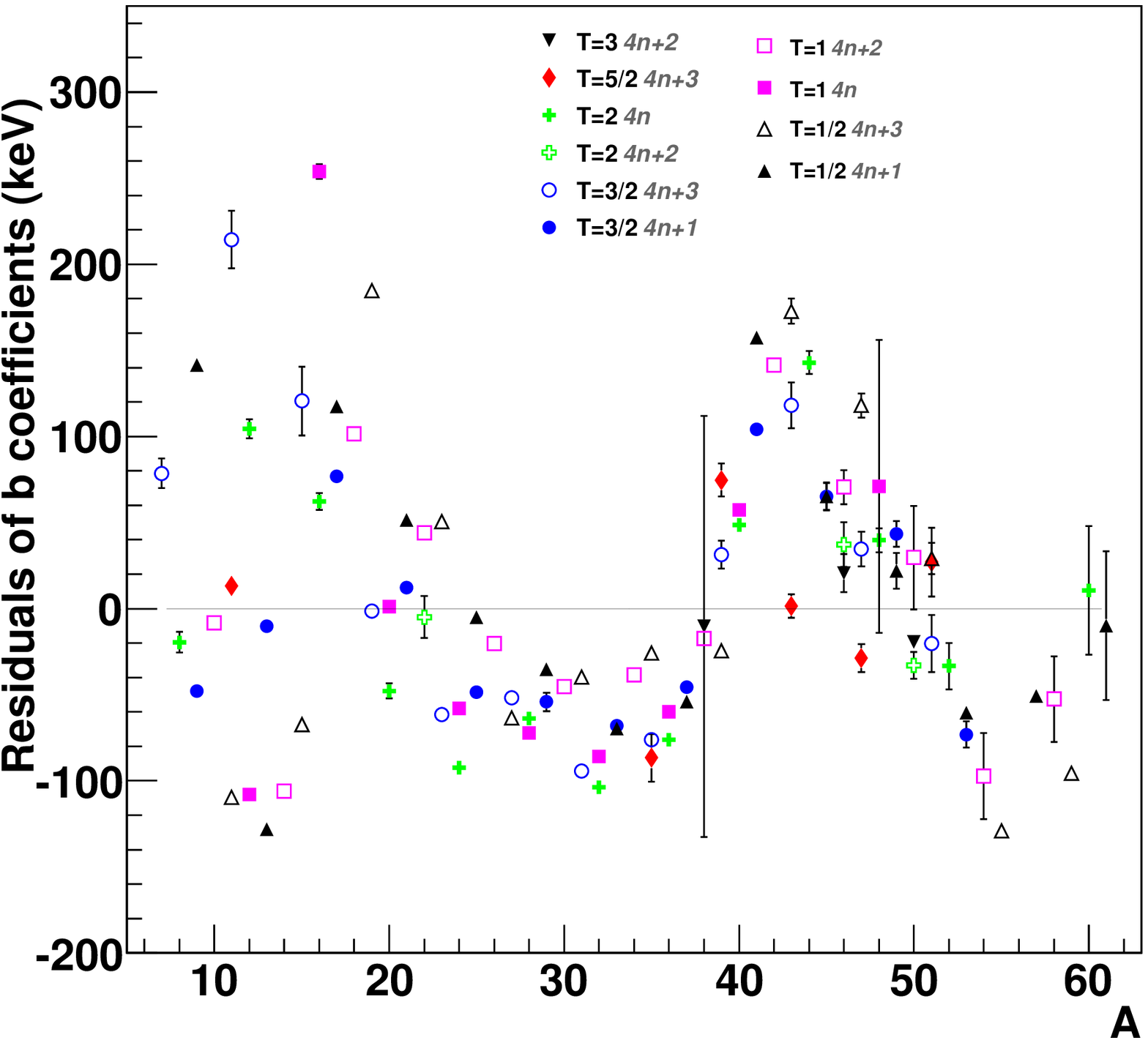}
\caption{Residual $b$ coefficient values separated into $(4n)$ groups follow similar patterns. The error bars are the IMME coefficient uncertainty, and reflect the original experimental data precision. The dispersion in the residual values dominates the overall precision of the analytical model.  
\label{fig:bFullResidualsNoE} }
\end{figure}
As initially suggested by Wigner and built upon over the years~\cite{2009LenziBently}, the $b$ coefficients should provide first-order effects of the Coulomb energy. In order to extract underlying symmetries we adopt the logic described by J{\"a}necke in a key paper~\cite{1966Janecke}, based on the assumption that the nucleus may be taken to be a homogeneously charged sphere. 

The Coulomb energy $E{_C}$ can then be written as 
\begin{equation}
E{_C} = \frac{3 e^2}{5R}Z(Z-1) \simeq \frac{3 e^2}{5 r_0 A^{1/3}}\left[\frac{A}{4}(A-2) + (1-A)T_z + \frac{1}{3}T{_z^2}\right]
\label{eqnIsovector}
\end{equation} 
Then, we can write
\begin{equation}
E{_C^{(0)}} = \frac{3 e^2 A(A-2)}{20 r_0 A^{1/3}}~~~~~E{_C^{(1)}} = -\frac{3 e^2 (A-1)}{5 r_0 A^{1/3}}~~~~~E{_C^{(2)}} = \frac{e^2}{5 r_0 A^{1/3}} 
\label{eqnSpherical}
\end{equation} 
and so we expect $b\simeq \Delta_{nH} - 0.720\times (A-1)/A{^{1/3}}$ taking $e^2=1.44$~MeV.fm and $r_0=1.2$~fm. Similar reasoning leads to $c\simeq 0.720\times A^{-1/3}$~MeV.

\subsection{Global $b$ function}
A simple averaged and unweighted $A^{2/3}$ linear fit returns a slope of $720$~keV, in agreement with the first order spherical nucleus model, and a global offset of $1868$~keV, of which $782$~keV is the n-$^1$H mass difference. Initially, the global straight line dependency of the $b$ coefficients was subtracted from the individual experimental data points to filter out the dominant term and reveal the underlying residual $b$ coefficient shape. However, a natural ``staggering" effect~\cite{YiHuaLam,1966Janecke} appears, as shown in figure~\ref{fig:unseparatedbStaggeringNoE}, which is larger for $T=1/2$ and $T=3/2$ multiplets. The oscillations between the two datasets also appear to be out phase. The effect is more pronounced for lower $A$, and gradually washes out beyond A$\sim$40.

The more complete $(A-1)/A^{1/3}$ spherical model gives a better description of the data, and is shown as a full line in figure~\ref{fig:bCoeffsAverageLineNoE}.  Separating the multiplets into 4-nucleon systems $(4n)$, removes the ``staggering" and makes it possible to establish more precise global fits to the data. The spherical description $b \simeq \Delta_{nH} - 720\times S{_b}\times(A-1)/A{^{1/3}} + C{_b}$, is fitted to the data, leaving $S_b$ and $C_b$ as free parameters. $S_b=1$ implies perfect agreement between the model and the data. The results are shown in table~\ref{tab:b-coeffSphericalFits} and the residual values, with respect to the individual data points, are shown in figure~\ref{fig:bFullResidualsNoE}. The global tendencies lead to $b$ residuals which are seen to follow smooth, similar shapes. A large dispersion in the residual values is observed at lower $A$, and this dispersion is similar for the different multiplets and $(4n)$ groups for $A>20$. The experimentally induced error bars on the IMME $b$ coefficients are drawn in figure~\ref{fig:bFullResidualsNoE}. However, the dispersion of the residual values with respect to the analytical function, dominates the precision, and varies as a function of $A$. The analytical coefficient values $S_b$ show that the spherical nucleus hypothesis is good to within 3\% for most multiplet groups, $T=1/2,(4n+3)$ being the exception. It is worth noting that the $T=3/2,(4n+3)$ and $T=2,(4n)$ multiplets are spherical to within 0.5\% and return nearly identical results.   
\begin{figure}[h] 
\centering
\includegraphics[height=11cm,clip]{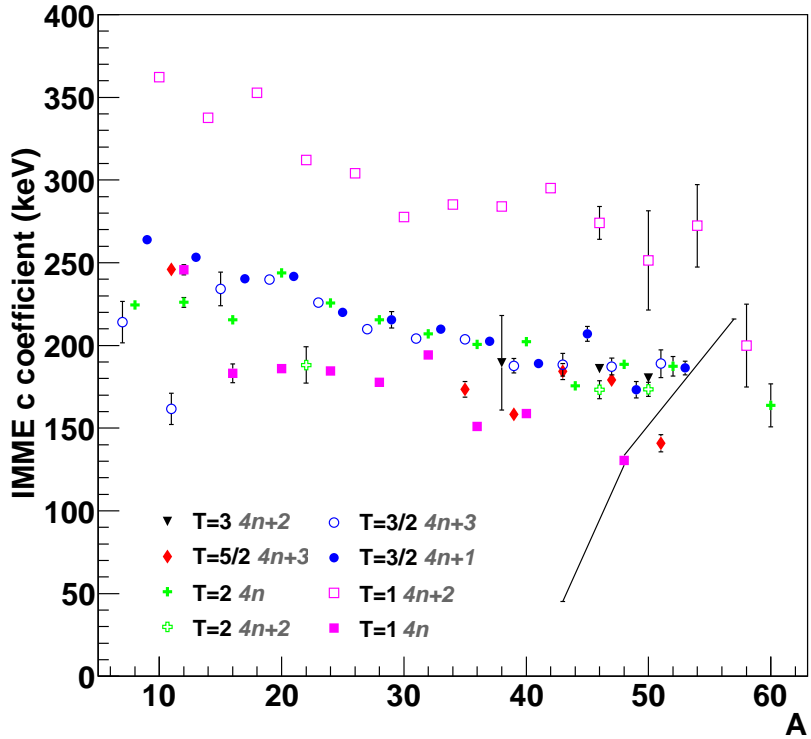}
\caption{IMME $c$ coefficients extracted from best fits. The large $\pm85$~keV error associated with the $T=1, A=48$ multiplet has been distorted to avoid overlapping with other data. \label{fig:NewcCoeffs4nGroupsvsANoE} }
\end{figure}  

Finally, using the results of each individual multiplet and of each $(4n)$ group, a global averaged function was established. The $T=3$ multiplet, which covers a very limited range in $A$, did not contribute to this part of the study. The final global function fixes $S_b= 0.979$ and $C_b = 702$~keV, as drawn in figure~\ref{fig:bCoeffsAverageLineNoE}, and may be used for rough calculations. 
\begin{table}[htb]
\footnotesize
\begin{center}
\begin{tabular}{llcc} 
\hline
Isospin ($T$) & $(4n)$ subgroup  &    $S_b$        &   $C_b$ (keV) \\
\hline
1/2 & $(4n+1)$    &  0.975 &  708 \\
1/2 & $(4n+3)$    &  0.958 &  446 \\
1   & $(4n)$      &  0.987 &  754 \\
1   & $(4n+2)$    &  0.971 &  601 \\
3/2 & $(4n+1)$    &  0.973 &  627 \\
3/2 & $(4n+3)$    &  0.993 &  845 \\
2   & $(4n)$      &  0.995 &  845 \\
2   & $(4n+2)$    &  0.975 &  676 \\
5/2 & $(4n+3)$    &  0.981 &  789 \\
3   & $(4n+2)$    &  1.016 &  1060 \\
\hline
\end{tabular}
\caption{\footnotesize Linear fits to the IMME $b$ coefficients based on the homogeneous spherical nucleus model, resulting in the analytical function $b \simeq \Delta_{nH} - 720\times S_b \times (A-1)/A{^{1/3}} + C_b$~keV.  \label{tab:b-coeffSphericalFits}}
\end{center}
\end{table}

\subsection{Global $c$ function}
The best fit IMME $c$ coefficients, which also show a natural ``staggering" effect for $T=1$ multiplets, were filtered into $(4n)$ groups and are shown in figure~\ref{fig:NewcCoeffs4nGroupsvsANoE}.
The $T=1,(4n)$ and $(4n+2)$ groups separate out, all groups following similar $A$-dependent, smoothly decreasing shapes. The homogeneously charged spherical model gives interesting results for this tensor term since, when fitted to the approximate $720*A^{-1/3}$~keV function, more than 93\% of the expected strength is observed for $T=1$ and in the limited $A$ range available for $T=3$. For the other cases, 58\% to 68\% of the strength is observed. 

\begin{figure}[h] 
\centering
\includegraphics[height=11cm,clip]{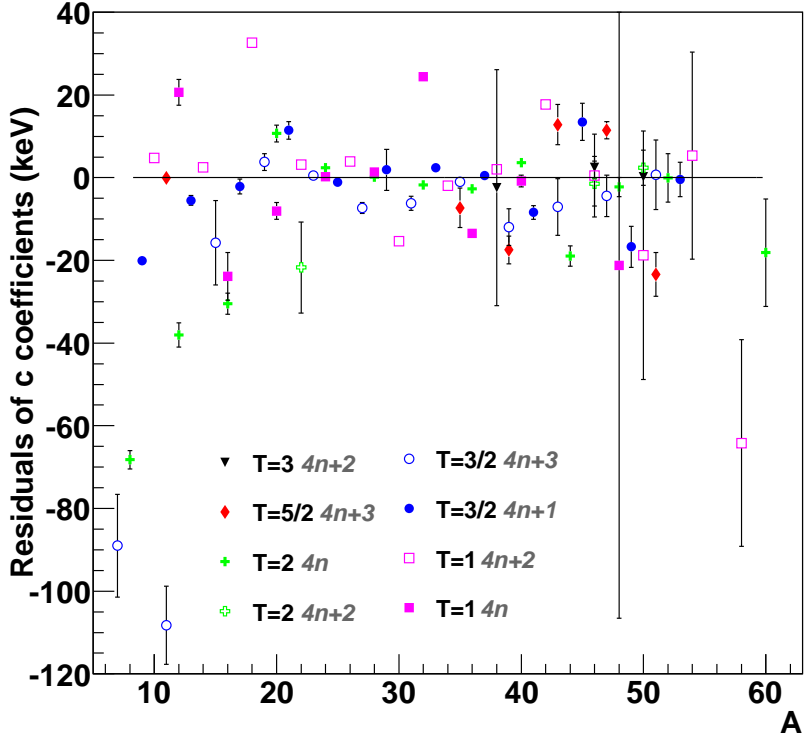}
\caption{$c$ coefficient residual values between individual IMME values and the global analytical $c$ function $c_{general} \simeq 3 \times (260\times 0.630/A{^{1/3}} + G)$~keV, based on the spherical model of the Coulomb energy tensor component. $G$ values, obtained for each multiplet and $(4n)$ group, are given in table~\ref{tab:c-coeffSphericalFits}. Better results can be obtained using the individual $c$ parametrizations with the values as listed in the columns $S_c$ and $C_c$ of the same table.
\label{fig:cCoeffResidualsAveragedFit} }
\end{figure}

Better fits to each multiplet are obtained from the model $c \simeq 3 \times (260~S{_c}/A{^{1/3}} + C{_c})$ with $S{_c}$ and $C{_c}$ as free parameters of the fit, and take the values listed in table~\ref{tab:c-coeffSphericalFits}. 

Global averaged, linear $A$-dependent functions, accurate to within $\pm$20~keV, are obtained by using the basic spherical description $3 \times (260\times 0.630/A{^{1/3}} + G)$~keV, where the only free parameter is $G$. The individual $G$ values are listed in the last column of table~\ref{tab:c-coeffSphericalFits}. Finally, a general, but crude estimate is given by $c_{general} \simeq 3 \times (260\times 0.630/A{^{1/3}} + 21.5)$~keV, shifted by $+85$~keV and $-42.5$~keV for $T=1,(4n+2)$ and $(4n)$ datasets, respectively.

\begin{table}[htb]
\footnotesize
\begin{center}
\begin{tabular}{llcc|| c} 
\hline
Isospin ($T$) & $(4n)$ subgroup  &    $S_c$       &   $C_c$ (keV) &   $G$ (keV)\\
\hline
1   & $(4n)$      &  0.946 & $-15.34$  &  9 \\
1   & $(4n+2)$    &  0.681 &  45.42  & 49 \\ 
3/2 & $(4n+1)$    &  0.583 &  25.92  & 22 \\
3/2 & $(4n+3)$    &  0.681 &  17.86  & 22 \\
2   & $(4n)$      &  0.627 &  21.96  & 22 \\
2   & $(4n+2)$    &  0.241 &  41.94  & 16 \\
5/2 & $(4n+3)$    &  0.615 &  15.49  & 14 \\
3   & $(4n+2)$    &  0.992 &  $-4.53$  & 19 \\
\hline
\end{tabular}
\caption{\footnotesize Linear fits to the IMME $c$ coefficients based on the homogeneous spherical nucleus model, leading to the analytical function $c \simeq 3 \times (260~S{_c}/A{^{1/3}} + C{_c})$, with $e^2=1440$~keV.fm, $r_0=1.2$~fm. $S_c$ and $C_c$ are free parameters of the fit. Fixing $S_c=0.630$, provides a general function $c_{general} \simeq 3 \times (260\times 0.630/A{^{1/3}} + G)$, good to within $\pm20$~keV. The $G$ values of the general function are listed in the last column, and the dispersion of the residual values are shown in figure~\ref{fig:cCoeffResidualsAveragedFit}.\label{tab:c-coeffSphericalFits}}
\end{center}
\end{table}

\subsection{IMME $|b/c|$ }
\begin{figure}[h]
\centering
\includegraphics[width=11cm,clip]{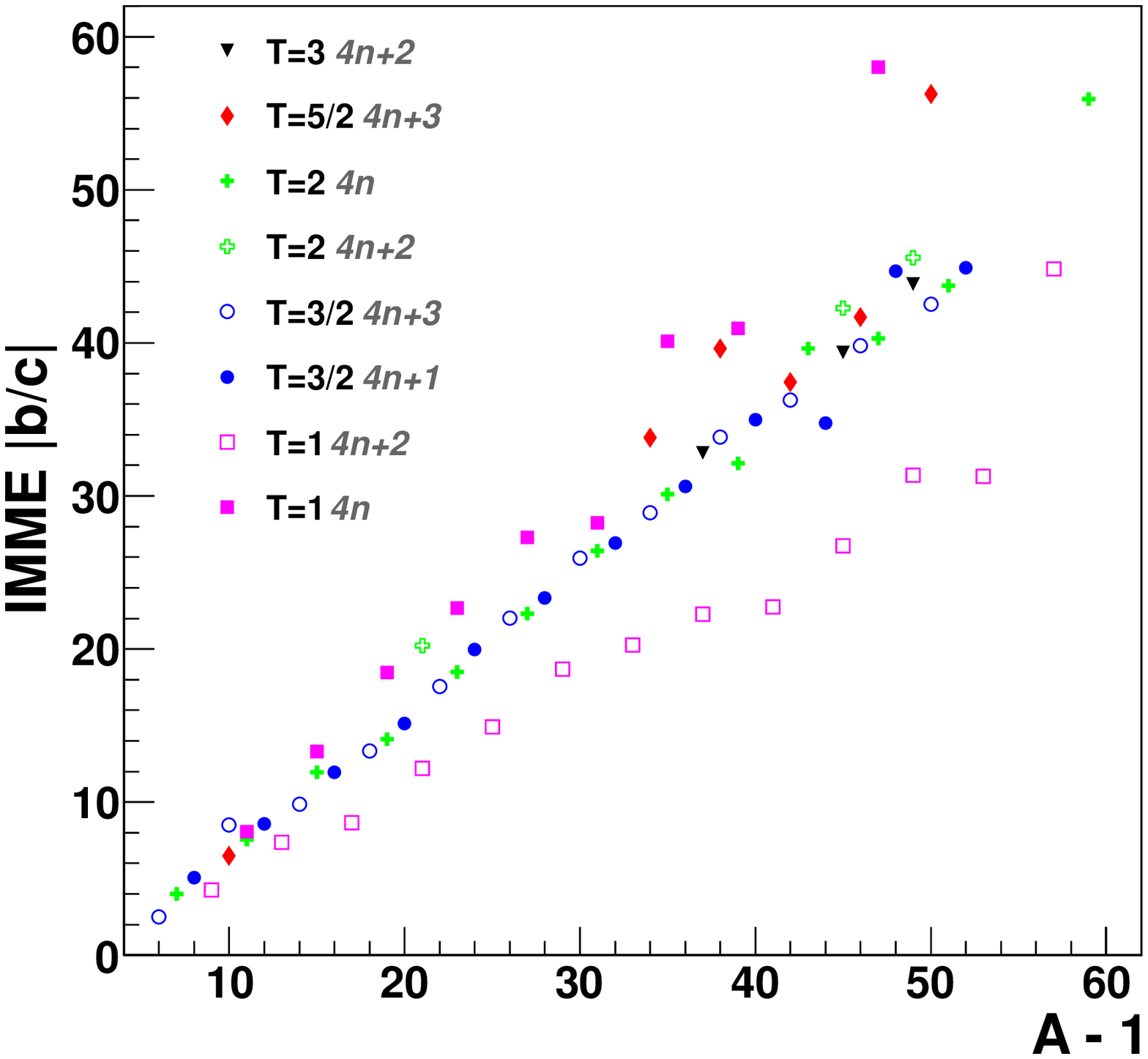} 
\caption{Ratio of $b$ to $c$ coefficients. The homogeneously charged sphere model predicts a linear dependancy as a function of $A-1$. There is agreement to first order, but there are clearly other underlying structures.  \label{fig:bOvercAm1NoE} }
\end{figure}

Within the confines of our simple model, the $|b$/$c|$ ratio is expected to evolve linearly as a function of $A-1$. This is seen to be roughly the case, however, as shown in figure~\ref{fig:bOvercAm1NoE}, there are clearly other underlying structures. The changes in slope are tied in to the extra constant terms found in the IMME fits. The coefficient ratio is roughly $|b/c| = 0.96*(A-1)-2.8$ with an extra $20\%$ on the slope for $T=1,(4n)$ and $20\%$ less for the $T=1,(4n+2)$ group, and is reasonable for crude calculations.

\subsection{IMME $d$ coefficients}
Non-zero cubic contributions are observed in very few cases after application of the $\pm$3$\sigma$ filter. The most probable set of coefficients for $A=9$ and $T=3/2$ involves a $d$ coefficient of 6.7$\pm$1.5~keV, in keeping with the results obtained nearly 40 years ago~\cite{1975Kashy}. The $A=35$ dataset, based on four high precision mass measurements, returns a $d$ coefficient of $-3.37\pm$0.38~keV, in agreement with previous work~\cite{A35Yazidjian2007} and merits closer inspection. The result for $A=53$, with a less precise value of $d = 40\pm10$~keV, would benefit from higher resolution measurements, especially in the determination of the $^{53}$Co$^i$ mass excess. 

In the $T=2$ dataset, the well documented case for $A=8$ returns $d = 9.42\pm1.88$~keV and is discussed in paragraph~\ref{Aeq9}. Finally, the exceptional $T=2, A=32$, as discussed in paragraph~\ref{Aeq32}, returns $d=0.89\pm0.11$~keV, in agreement with previous reports~\cite{A32Kwiatkowski2009}, and is as yet unexplained. This most intriguing case requires further experimental investigation.

\subsection{Mass Predictions}\label{predict}
Using the unrounded indiviual IMME coefficients as listed in~\ref{IMMEtables}, unmeasured masses can be estimated to tens of keV precision. The $T=3/2$ multiplet mass excess estimates are all $T_z=-1/2$ IAS and are: $^{15}$O$^i$~=~$14029\pm10$~keV, $^{39}$Ca$^i$~=~$-20905\pm10$~keV, $^{43}$Ti$^i$~=~$-25126\pm10$~keV, $^{47}$Cr$^i$~=~$-30400\pm50$~keV, $^{51}$Fe$^i$~=~$-35760\pm50$~keV. The $T=2$ $^{16}$F$^i$ missing IAS is estimated have a mass excess of $20773\pm10$~keV.  The errors are estimated from a detailed study of the capacity of the IMME to predict known masses, as described in~\ref{predictability}. As a rough guide, for excited IAS up to $A=44$ the predictions are accurate to $\pm10$~keV, and $50$~keV thereafter.

The IMME $b$ and $c$ coefficients can also be estimated from global trends in the $b$ and $c$ coefficients. This is useful when there is not enough data available for an individual IMME fit, for example, the $T=1$, $A=52$ and $A=56$ multiplets.  However, estimating the $a$ coefficient, a term which is not only independent of $T_z$ but is also the dominating term, is more problematic. For integer valued multiplets the $N=Z$ IAS is exactly the $a$ coefficient value and is consequently a valuable source of information. However, for multiplets of isospin greater than $T=1$, the missing data concerns relatively high excitation levels where fragmentation due to isospin mixing is more probable. Experimentally, these states are expected to be very short-lived particle unbound states and may remain inaccessible.  

Nonetheless, there are some interesting cases, discussed in the following paragraphs, that can be estimated using the global trends of the coefficients.

\subsubsection{$T{_z}=-1, ^{44}$V ground state}
From the $T=1,(4n)$ multiplet description one would expect a primary $b$-value of $\sim-7120$~keV associated with a $b$ residual estimated at $\sim100 - 160$~keV leading to $b\sim-6960$~keV to $-7020$~keV. The $c$ coefficient is estimated from the average of the $A=40$ and $48,(4n)$ values at $c\sim145$~keV, leading to an estimated mass excess of $-23777$~keV to $-23837$~keV, some $150$~keV to $200$~keV higher than the experimental result of $-23980+80-380$~keV for this isomer contaminated odd-odd nucleus.  

\subsubsection{$T{_z}=-1, ^{52}$Co$^i$}
The straight line of the $T=2, |b/c|$ ratio, as shown in figure~\ref{fig:bOvercAm1NoE}, can also be used to identify outlying cases. Excluding $^{52}$Co$^i$ from the IMME fit returns a $|b/c|$ value that lies along the $T=2$ diagonal.  Using the IMME results, as given in table~\ref{tab:Teq2coeffs}, the mass excess for $^{52}$Co$^i$ is expected to be $-31590\pm50$~keV, $26$~keV lower than, but in good agreement with, the current {\sc AME2012} value. 

An estimation based on the global tendencies is less precise and in this case the $T=2, A=52$ coefficients are estimated at $a =\Delta M(^{52}Fe^j)$, $b=-8162$~keV with negligible residual, and $c= 197$~keV, leading to a mass excess of $-31416$~keV, some $150$~keV above the current {\sc Ame2012} value. 

\section{Summary}

Experimental data, accumulated over more than 50 years of fundamental research, have been organised into isobaric multiplets and presented for T=1 to T=4 datasets. The T=1/2 mirror ground states are as found in {\sc Ame2012}. A few particular cases have been corrected and updated with respect to {\sc Ame2012} and {\sc Nubase2012}. The spin-parity attributions of the multiplets are unambiguous and, in the absence of isospin mixing, many previously estimated spin-parities can now be fixed on the basis of IAS membership. IMME coefficients have been extracted whenever possible and fragmented states, requiring further study, have not been included in the IMME tables. The IMME coefficients, from quadratic and higher degree polynomial fits are given, as unrounded values, for use in calculations for $T=1/2$ to $T=3$ multiplets. Higher order IMME contributions are observed for $A=8$ and $A=9$, where a better identification of possibly fragmented states would clarify these cases. The cubic contributions for $A=32$ and $A=35$ are confirmed with this evaluated and adjusted dataset. 
IMME coefficients for individual isobaric sets have been used to predict currently unmeasured masses. Key nuclides, allowing to complete the IAS multiplets, have been highlighted throughout the text. Global coefficient tendencies have also been extracted allowing to estimate the coefficients of unmeasured IAS multiplets, and may be used for rough calculations. It is hoped that this work will be useful in both theoretical and experimental work.

\section{Acknowledgements}

We are greatly indebted to A.H.~Wapstra who, throughout his lifetime, kept detailed notes on IAS and which were the starting point for this work. One of us thanks Piet Van Isacker from GANIL, in France, for his many illuminating discussions that were made possible thanks to the generous invitation of the Kavli Institute for Theoretical Physics, CAS, Beijing, China during April 2013.

\normalsize
\clearpage

\begin{appendix}
\section{Special IAS Notation}
The IAS states studied here are all analogue ground states given that each multiplet is delimited by ground state nuclei. In order to differentiate between ground and excited states new notation has been introduced in the {\sc Ame}. The $A=36, T=1$ and $T=2$ IAS multiplets are used to illustrate the new naming convention. The lowest lying excited IAS are labelled with a superscript $i$ and the higher IAS levels with $j$. The $T=1$ triplet is then $^{36}$Cl, $^{36}$Ar$^i$ and $^{36}$K, whereas the $T=2$ quintuplet is composed of  $^{36}$S, $^{36}$Cl$^i$, $^{36}$Ar$^j$, $^{36}$K$^i$ and $^{36}$Ca. 

\section{Isomeric Exceptions}
Some IAS are also isomeric states, and this has led to some difficulties in having a unified and simple notation system. In the {\sc Ame} successively higher isomeric states are labelled with $m$ and $n$ superscripts. Although this work focuses on IAS, the isomeric states are of a more fundamental nature and so the isomer labels are used in preference to the IAS labelling scheme. The three cases affected by this are $^{16}$N$^{i}$, $^{26}$Al$^{i}$ and $^{38}$K$^{i}$ and are each labelled with an $m$ superscript.

\section{$N=Z, T=1$ ground state Exceptions}
These ground states are exceptional since they do not follow the general convention that the lowest lying energy level is $T=0$, but have $T=1$. In these particular cases it is the $T=1$ ground state which is the IAS. However, following standard conventions, the ground state has no particular label. There are 7 cases affected by this they are $^{34}$Cl, $^{42}$Sc, $^{46}$V, $^{50}$Mn, $^{54}$Co, $^{62}$Ga and $^{70}$Br, they are all J$^{\pi}$ 0$^+$ states, and in the IAS notation would all carry an $i$ superscript.  

\section{Changes with respect to {\sc Ame2012} and {\sc Nubase2012}}
\noindent {\bf Data withdrawn}\newline
Two $T=3/2$ IAS were mistakenly included in {\sc Nubase2012}. They are $^{43}$Ti$^i$ and $^{47}$Cr$^i$ for which no original data source could be found. $^{19}$Ne${^i}$ has been corrected. No data source could be located for the {\sc Nubase2012} $^{44}$Ca$^i$ $T=3$ state, and it is removed.

\vskip0.5cm
\noindent {\bf Data added}\newline
The mass excess of $^{31}$S$^i$, a member of the $T=3/2$ multiplet, has been updated with respect to {\sc Nubase2012}. The updated value leads to an excitation energy of 6280.60$\pm$0.16~keV. A recent measurement~\cite{2013Tripathi} of the fragmented $^{55}$Ni$^{i}$ $T=3/2$ state has been included.  

$^{12}$N$^i$ is updated using reference~\cite{2012Jager11}, $^{11}$B$^j$ is from reference~\cite{2012Charity40}, and there is supplementary $^{56}$Cu$^i$ data from~\cite{2012Orrigo}. $^{19}$Ne${^i}$ data from~\cite{1998Utku} has been included, and the spin-parity attribution tentatively assigned in~\cite{2007Nesaraja} is confirmed here.\newline

\vskip0.5cm
\noindent {\bf Notation}\newline
Some states were incorrectly labelled in {\sc Ame2012} and {\sc Nubase2012} and have been corrected in this paper and are summarized in table~\ref{tab:AMEerrata}. The correct labels, as they should have been attributed, are given in column~2. It is underlined that the associated numerical data is not affected by these changes. 
\small{
\begin{table}[h]
\begin{center}
\begin{tabular}{ccl} 
\hline
{\sc Ame2012}    &  Corrected Label  &  Multiplet \\
\hline
$^{16}$N$^{j}$      &        $^{16}$N$^{i}$        &    $T=2$ IAS of $^{16}$C.          \\
$^{28}$Si$^{i}$     &        $^{28}$Si$^{j}$       &    $T=2$ IAS of $^{28}$Mg.   \\
$^{28}$Si$^{j}$     &        $^{28}$Si$^{i}$       &    $T=1$ IAS of $^{28}$Al. \\
$^{34}$Ar$^i$       &                              &    $T=2$ state, not $T=3$. \\
$^{42}$Sc$^{i}$     &   $^{42}$Sc$^{m}$ or  $^{42}$Sc$^{n}$    &   probably an isomer.    \\
\hline
\end{tabular}
\caption{Corrected IAS labels. \label{tab:AMEerrata}} 
\end{center}
\end{table}
}

\normalsize
\section{IMME predictability}\label{predictability}

The capacity of the IMME to predict masses was evaluated using the subset of T=3/2 and T=2 experimentally complete multiplets.
Five out of six of the currently unmeasured masses are $T=3/2$, $T_{z}=-1/2$ cases, and so this particular configuration is illustrated here.

\begin{figure}[h]
\centering
\includegraphics[width=13cm,clip]{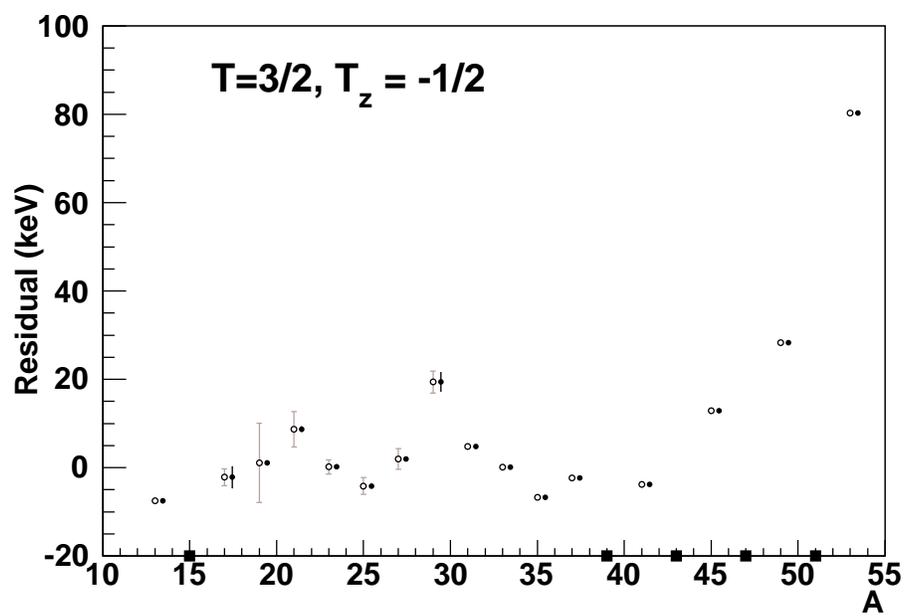} 
\caption{Residual values for $T=3/2$, $T_{z}=-1/2$ measurements and IMME predictions, based on experimentally complete multiplets. The residual value, shown in keV as a function of atomic mass number $A$, is the experimental mass excess value minus the IMME prediction. The atomic mass number of currently unmeasured IAS are highlighted by a small full squares placed on the x-axis.  \label{fig:predictability} }
\end{figure}

The $T_{z}=-1/2$ component of fully complete multiplets was removed from the dataset. The three remaining data points were used to predict the artificially missing value. The predicted and the experimentally determined results were then compared. The residual value, constructed from the difference between observation and prediction, are shown in figure~\ref{fig:predictability}. The error bars of the original experimental uncertainties is shown, along with the IMME prediction uncertainty, slightly displaced to the right, as generated by the uncertainty in the IMME coefficients. Most of the IMME predictions have sub-keV errors, and show up as a small dot on the graphics, although the $A=17$ and $A=29$ predictions give errors of the same magnitude as the experimental observation. The effect that clearly dominates the predictive capacity is the dispersion of the residuals, as opposed to the experimental or prediction uncertainties themselves. This is similar to what has also been observed for the $b$ and $c$ coefficient residuals, as shown in figures~\ref{fig:bFullResidualsNoE} and~\ref{fig:cCoeffResidualsAveragedFit}. 

Consequently, in order to take into account the dispersion of the residual values, a vaue of $\pm10$~keV uncertainty is adopted for $A<40$ and $\pm50$~keV above. 

\clearpage
\section{IAS adjusted experimental masses}\label{IAStables}
\noindent {\bf Explanation of tables} \newline
\begin{table}[ht]
\small
\begin{tabular}{lll} 
  Nuclide      && Nuclidic name: mass number  $A =   N + Z$ and element symbol. \\
               && Elements with upper suffix `$i$', `$j$' are IAS assignments
\\\\
  Mass excess  && Mass excess [$M$(in u)$-A$], in keV, and its one standard \\
               && deviation uncertainty as given in {\sc Ame2012}~\cite{AME2012b}. \\
               && Rounding-off policy: in cases where the furthest-left significant \\
               && digit in the error is larger than 3, values and errors are \\
               && rounded-off, but not to more than tens of keV.\\
               && (For example: $2345.67\pm2.78 \to 2345.7\pm2.8$,\\
               && $2345.67\pm4.68 \to 2346\pm5$, but $2346.7\pm468.2 \to 2350\pm470)$.
\\\\
  Excitation energy
                && Energy difference, in keV, between levels adopted as higher levels \\
                && and the ground state (gs), accompanied by its one standard deviation\\
                && uncertainty, as given in {\sc Ame2012} when derived from the \\
                && {\sc Ame}, otherwise as given by {\sc Ensdf}. \\
                && The rounding-off policy is the same as for the mass excesses. 
\\\\
  $J^{\pi}$  && Spin and parity:   \\
               &  () & uncertain spin and/or parity.\\
               & \#   & Deduced, not measured.   
\\\\
  $T_{z}$    &&  Isospin projection of isobaric analogue states (IAS).\\
             & frg & Experimentally observed fragmented levels.
\\\\
Data Source   &  MD    & mass doublet    \\
              &  RQ    & reaction $Q$-value  \\
              &  p, 2p & one-, two-proton decay \\
              &  Two-digit & Year of {\sc Ensdf} file archive\\
              & number & 
\end{tabular}
\end{table}

\footnotesize{
\hspace{-1.9cm}{\em Remarks.~}All ground state masses are from {\sc Ame2012}. The source of the associated spin-parity data is given as a two-digit number in the Data Source column. \newline 
Excited states are either from {\sc Ensdf} or from reaction data. In most cases the reaction data $Q$-value also exists in the {\sc Ensdf} archives.\newline
Spin-parties are from {\sc Ensdf} or deduced from IAS multiplets. }

\normalsize
\clearpage
\LTright=0pt
\LTleft=0pt
\footnotesize

\begin{longtable}{
 l@{\extracolsep{0pt}}                               
 l@{\extracolsep{\fill}}                             
 *{4}{r@{\extracolsep{0pt}}l@{\extracolsep{\fill}}}  
 r@{\extracolsep{\fill}}                             %
 r@{\extracolsep{\fill}}                             %
 r@{\extracolsep{\fill}}                             %
 c@{\extracolsep{\fill}}                             
 c@{\extracolsep{\fill}}                             
 c@{\extracolsep{\fill}}                             
 r                                                   
 }
\caption{Adjusted IAS masses of $T=1$ multiplets.\newline\\ \label{tab:J1}} \\[-6pt]
\hline 
&Nuclide&\multicolumn4c{Mass~Excess}&\multicolumn4l{Excitation}& \multicolumn1c{~~~~~$J^{\pi}$} & \multicolumn1c{~~~$T_z$} & \multicolumn1c{Data}&\\
\noalign{\vskip3pt}
&&\multicolumn4c{(keV)}&\multicolumn4c{Energy~(keV)}& & &\multicolumn1c{Source}\\
\hline 
\endfirsthead
\multicolumn{3}{@{}c@{}}{\bf  Table~\ref{tab:J1}  cont'd\dots}\\[6pt]
\hline\noalign{\vskip3pt}
&Nuclide&\multicolumn4c{Mass~Excess}&\multicolumn4l{Excitation}& \multicolumn1c{~~~~~$J^{\pi}$} & \multicolumn1c{~~~$T_z$} & \multicolumn1c{Data}&\\
\noalign{\vskip3pt}
&&\multicolumn4c{(keV)}&\multicolumn4c{Energy~(keV)}& & & \multicolumn1c{Source}&\\
\hline 
\endhead
\hline
\noalign{\vskip3pt}
\multicolumn{13}{r}{\itshape $T=1$ cont'd\dots}\\
\endfoot
\noalign{\vskip3pt}\hline
\endlastfoot
     & $^{6}$He        &$  17592    $&$.09   $&$    0    $&$.05  ~$&$          $&$     ~$&$        $&$     ~$&$ 0^+$  & 1  & 02 &\\
     & $^{6}$Li$^i$    &$  17649    $&$.76   $&$    0    $&$.10  ~$&$  3562    $&$.88  ~$&$   0    $&$.10  ~$&$ 0^+$  & 0  & 02 &\\
     & $^{6}$Be        &$  18375    $&$      $&$    5    $&$     ~$&$          $&$     ~$&$        $&$     ~$&$ 0^+$  & -1 & 02 &
 \\\\
     & $^{8}$Li        &$  20945    $&$.80   $&$    0    $&$.05  ~$&$          $&$     ~$&$        $&$     ~$&$ 2^+ $ & 1  & 05 &\\
     & $^{8}$Be$^i$    &$  21568    $&$      $&$    3    $&$     ~$&$ 16626    $&$     ~$&$   3    $&$     ~$&$ 2^+$ 
     & frg  & &\footnote{strongest frg; other: 296(3) higher I(16626)/I(16922)=1.22 in 6Li(6Li,a) and 1.15 in 10B(d,a); see~\cite{2004Tilley06} p.213.}\\
     & $^{8}$B         &$  22921    $&$.6    $&$    1    $&$.0   ~$&$          $&$     ~$&$        $&$     ~$&$ 2^+$  & -1 & 05 &
\\\\
\end{longtable}


\clearpage
\LTright=0pt
\LTleft=0pt
\footnotesize
\begin{longtable}{
 l@{\extracolsep{0pt}}                               
 l@{\extracolsep{\fill}}                             
 *{4}{r@{\extracolsep{0pt}}l@{\extracolsep{\fill}}}  
 r@{\extracolsep{\fill}}                             %
 r@{\extracolsep{\fill}}                             %
 r@{\extracolsep{\fill}}                             %
 c@{\extracolsep{\fill}}                             
 c@{\extracolsep{\fill}}                             
 c@{\extracolsep{\fill}}                             
 r                                                   
 }
\caption{Adjusted IAS masses of $T=3/2$ multiplets.\newline\\ \label{tab:J2}} \\[-6pt]
\hline 
&Nuclide&\multicolumn4c{Mass~Excess}&\multicolumn4l{Excitation}& \multicolumn1c{~~~~~$J^{\pi}$} & \multicolumn1c{~~~$T_z$} & \multicolumn1c{Data}&\\
\noalign{\vskip3pt}
&&\multicolumn4c{(keV)}&\multicolumn4c{Energy~(keV)}& & &\multicolumn1c{Source}\\
\hline 
\endfirsthead
\multicolumn{3}{@{}c@{}}{\bf  Table~\ref{tab:J2}  cont'd\dots}\\[6pt]
&Nuclide&\multicolumn4c{Mass~Excess}&\multicolumn4l{Excitation}& \multicolumn1c{~~~~~$J^{\pi}$} & \multicolumn1c{~~~$T_z$} & \multicolumn1c{Data}&\\
\noalign{\vskip3pt}
&&\multicolumn4c{(keV)}&\multicolumn4c{Energy~(keV)}& & & \multicolumn1c{Source}&\\
\hline 
\endhead
\hline
\noalign{\vskip3pt}
\multicolumn{13}{r}{\itshape $T=3/2$ cont'd\dots}\\
\endfoot
\noalign{\vskip3pt}\hline
\endlastfoot
     & $^{7}$He        &$  26073    $&$ $&$    8    $&$ $&$          $&$ $&$        $&$ $&$ (3/2)^-$  &  3/2 &  03 &\\
     & $^{7}$Li$^i$    &$  26150    $&$ $&$   30    $&$ $&$  11250   $&$ $&$   30   $&$ $&$ 3/2^-$    & 1/2  & RQ 03&\\
     & $^{7}$Be$^i$    &$  26750    $&$ $&$   30    $&$ $&$  10980   $&$ $&$   30   $&$ $&$ 3/2^-$    & -1/2 & RQ 03 &\\
     & $^{7}$B         &$  27677    $&$ $&$   25    $&$ $&$          $&$ $&$        $&$ $&$ (3/2)^-$  & -3/2 & 03 &
\\\\
     & $^{9}$Li        &$  24954  $&$.90 $&$ 0$&$.19 $&$          $&$ $&$        $&$ $&$ 3/2^-$  &  3/2 &  06 &\\
     & $^{9}$Be$^i$    &$ 25738   $&$.9  $&$ 1$&$.8  $&$ 14390$   &$.5 $&$     1$&$.8$& $3/2^-$ & 1/2  & RQ 06 &\footnote{possibly isospin mixed~\cite{2012Brodeur}. }\\
     & $^{9}$B$^i$     & $27071$&$.1 $&$     2$&$.3  $&$ 14654$&$.7  $&$   2$&$.5 $& $3/2^-$ &-1/2& RQ 06 &\footnote{possibly isospin mixed~\cite{2012Brodeur}. } \\
\end{longtable}


\clearpage
\LTright=0pt
\LTleft=0pt
\footnotesize
\begin{longtable}{
 l@{\extracolsep{0pt}}                               
 l@{\extracolsep{\fill}}                             
 *{4}{r@{\extracolsep{0pt}}l@{\extracolsep{\fill}}}  
 r@{\extracolsep{\fill}}                             %
 r@{\extracolsep{\fill}}                             %
 r@{\extracolsep{\fill}}                             %
 c@{\extracolsep{\fill}}                             
 c@{\extracolsep{\fill}}                             
 c@{\extracolsep{\fill}}                             
 r                                                   
 } 
\caption{Adjusted IAS masses of $T=2$ multiplets.\newline\\ \label{tab:J3}} \\[-6pt]
\hline
&Nuclide&\multicolumn4c{Mass~Excess}&\multicolumn4l{Excitation}& \multicolumn1c{~~~~~$J^{\pi}$} & \multicolumn1c{~~~$T_z$} & \multicolumn1c{Data}&\\
\noalign{\vskip3pt}
&&\multicolumn4c{(keV)}&\multicolumn4c{Energy~(keV)}& & &\multicolumn1c{Source}\\
\hline 
\endfirsthead
\multicolumn{3}{@{}c@{}}{\bf  Table~\ref{tab:J3}  cont'd\dots}\\[6pt]
\hline\noalign{\vskip3pt}
&Nuclide&\multicolumn4c{Mass~Excess}&\multicolumn4l{Excitation}& \multicolumn1c{~~~~~$J^{\pi}$} & \multicolumn1c{~~~$T_z$} & \multicolumn1c{Data}&\\
\noalign{\vskip3pt}
&&\multicolumn4c{(keV)}&\multicolumn4c{Energy~(keV)}& & & \multicolumn1c{Source}&\\
\hline 
\endhead
\hline
\noalign{\vskip3pt}
\multicolumn{13}{r}{\itshape $T=2$ cont'd\dots}\\
\endfoot
\noalign{\vskip3pt}\hline
\endlastfoot
 & $^{8}$He  &   31609&.68 &    0&.09 &    &&&&                         0+   & 2&        05 &\\
 & $^{8}$Li$^i$ &   31768&&        5&&     10822&&       5&&               0+ & 1 &  RQ 05 &\\
 & $^{8}$Be$^j$ &   32436&.0   &   2&.0&   27494&.3&     2&.0 &            0+ & 0 &  RQ 05 &\\
 & $^{8}$B$^i$ &   33546&&        8 &&    10624 &&      8 &&              0+ & -1 &   RQ 05 &\\ 
 & $^{8}$C   &   35064&&       18 &&       &&&&                         0+ & -2 &      05&
\\\\
 & $^{12}$Be  &  25077&.8&      1&.9 &     &&&&                         0+ &  2 &        00 &\\
 & $^{12}$B$^i$ &  26088 &&      19&&     12719&&      19 &&              0+ &  1 &     RQ 00 & \\
\end{longtable}
\clearpage

\LTright=0pt
\LTleft=0pt
\footnotesize
\begin{longtable}{
 l@{\extracolsep{0pt}}                               
 l@{\extracolsep{\fill}}                             
 *{4}{r@{\extracolsep{0pt}}l@{\extracolsep{\fill}}}  
 r@{\extracolsep{\fill}}                             %
 r@{\extracolsep{\fill}}                             %
 r@{\extracolsep{\fill}}                             %
 c@{\extracolsep{\fill}}                             
 c@{\extracolsep{\fill}}                             
 c@{\extracolsep{\fill}}                             
 r                                                   
 }
\caption{Adjusted IAS masses of $T=5/2$ multiplets.\newline\\ \label{tab:J4}} \\[-6pt]
\hline 
&Nuclide&\multicolumn4c{Mass~Excess}&\multicolumn4l{Excitation}& \multicolumn1c{~~~~~$J^{\pi}$} & \multicolumn1c{~~~$T_z$} & \multicolumn1c{Data}&\\
\noalign{\vskip3pt}
&&\multicolumn4c{(keV)}&\multicolumn4c{Energy~(keV)}& & &\multicolumn1c{Source}\\
\hline 
\endfirsthead
\multicolumn{3}{@{}c@{}}{\bf  Table~\ref{tab:J4}  cont'd\dots}\\[6pt]
\hline\noalign{\vskip3pt}
&Nuclide&\multicolumn4c{Mass~Excess}&\multicolumn4l{Excitation}& \multicolumn1c{~~~~~$J^{\pi}$} & \multicolumn1c{~~~$T_z$} & \multicolumn1c{Data}&\\
\noalign{\vskip3pt}
&&\multicolumn4c{(keV)}&\multicolumn4c{Energy~(keV)}& & & \multicolumn1c{Source}&\\
\hline 
\endhead
\hline
\noalign{\vskip3pt}
\multicolumn{13}{r}{\itshape $T=5/2$ cont'd\dots}\\
\endfoot
\noalign{\vskip3pt}\hline
\endlastfoot
 & $^{11}$Li  &  40728&.3 &     0&.6 &    &&&&                         3/2- &  5/2  &     12 &\\
 & $^{11}$Be$^i$ &  41337&.2 &    20&.0 &  21160&&      20&&          (3/2-)\# &  3/2  & RQ~~~~  & \footnote{\cite{1997Teranishi07}. Addition to {\sc Ame2012}.}\\
 & $^{11}$B$^j$ & 42438 &&      80 &&    33570 &&     80  &&         (3/2-)\# &  1/2  & RQ~~~~ & \footnote{Update from~\cite{2012Charity40}.}
\\\\
 & $^{23}$F   &   3310&&       50&&      &&&&                     5/2+   &  5/2 &  07 &\\
 & $^{23}$Al$^i$ &  18530&&       60&&     11780&&      60&&        (5/2)+  & -3/2 &  p 07 &
\\\\
 & $^{27}$Na  &  -5518 &&       4&&          &&&&                 5/2+  &  5/2     &  11 &\\
\end{longtable}

\newpage
\LTright=0pt
\LTleft=0pt
\footnotesize
\begin{longtable}{
 l@{\extracolsep{0pt}}                               
 l@{\extracolsep{\fill}}                             
 *{4}{r@{\extracolsep{0pt}}l@{\extracolsep{\fill}}}  
 r@{\extracolsep{\fill}}                             %
 r@{\extracolsep{\fill}}                             %
 r@{\extracolsep{\fill}}                             %
 c@{\extracolsep{\fill}}                             
 c@{\extracolsep{\fill}}                             
 c@{\extracolsep{\fill}}                             
 r                                                   
 }
\caption{Adjusted IAS masses of $T=3$ multiplets.\newline\\ \label{tab:J5}} \\[-6pt]
\hline 
&Nuclide&\multicolumn4c{Mass~Excess}&\multicolumn4l{Excitation}& \multicolumn1c{~~~~~$J^{\pi}$} & \multicolumn1c{~~~$T_z$} & \multicolumn1c{Data}&\\
\noalign{\vskip3pt}
&&\multicolumn4c{(keV)}&\multicolumn4c{Energy~(keV)}& & &\multicolumn1c{Source}\\
\hline 
\endfirsthead
\multicolumn{3}{@{}c@{}}{\bf  Table~\ref{tab:J5}  cont'd\dots}\\[6pt]
\hline\noalign{\vskip3pt}
&Nuclide&\multicolumn4c{Mass~Excess}&\multicolumn4l{Excitation}& \multicolumn1c{~~~~~$J^{\pi}$} & \multicolumn1c{~~~$T_z$} & \multicolumn1c{Data}&\\
\noalign{\vskip3pt}
&&\multicolumn4c{(keV)}&\multicolumn4c{Energy~(keV)}& & & \multicolumn1c{Source}&\\
\hline 
\endhead
\hline
\noalign{\vskip3pt}
\multicolumn{13}{r}{\itshape $T=3$ cont'd\dots}\\
\endfoot
\noalign{\vskip3pt}\hline
\endlastfoot
 & $^{38}$S   & -26861 &&       7 &&      &&&&                         0+ &   3 &    08 &\\ 
 & $^{38}$Cl$^i$ & -21590 &&      24 &&     8208 &&     24 &&             0+ &   2 &  RQ 08 &\\
 & $^{38}$Ar$^j$ & -15940 &&      30 &&   18780  &&    30  &&            0+  &  1  & RQ 08 &
\\\\
 & $^{42}$Ar  & -34423&&        6 &&        &&&&                       0+   &    3 &    01 &\\
 & $^{42}$K$^i$ & -28570&&      100 &&     6450 &&    100 &&            (0+)\# &   2 &  01 &
\\\\
 & $^{46}$Ca  & -43138&.4 &     2&.3&     &&&&                         0+  & 3     &    00 &\\
\end{longtable}

\vskip1.0cm
\LTright=0pt
\LTleft=0pt
\footnotesize
\begin{longtable}{
 l@{\extracolsep{0pt}}                               
 l@{\extracolsep{\fill}}                             
 *{4}{r@{\extracolsep{0pt}}l@{\extracolsep{\fill}}}  
 r@{\extracolsep{\fill}}                             %
 r@{\extracolsep{\fill}}                             %
 r@{\extracolsep{\fill}}                             %
 c@{\extracolsep{\fill}}                             
 c@{\extracolsep{\fill}}                             
 c@{\extracolsep{\fill}}                             
 r                                                   
 }
\caption{Adjusted IAS masses of $T=4$ multiplets.\newline\\ \label{tab:J6}} \\[-6pt]
\hline
&Nuclide&\multicolumn4c{Mass~Excess}&\multicolumn4l{Excitation}& \multicolumn1c{~~~~~$J^{\pi}$} & \multicolumn1c{~~~$T_z$} & \multicolumn1c{Data}&\\
\hline 
\noalign{\vskip3pt}
&&\multicolumn4c{(keV)}&\multicolumn4c{Energy~(keV)}& & &\multicolumn1c{Source}\\
\hline 
\endfirsthead
\multicolumn{3}{@{}c@{}}{\bf  Table~\ref{tab:J6}  cont'd\dots}\\[6pt]
\hline\noalign{\vskip3pt}
&Nuclide&\multicolumn4c{Mass~Excess}&\multicolumn4l{Excitation}& \multicolumn1c{~~~~~$J^{\pi}$} & \multicolumn1c{~~~$T_z$} & \multicolumn1c{Data}&\\
\noalign{\vskip3pt}
&&\multicolumn4c{(keV)}&\multicolumn4c{Energy~(keV)}& & & \multicolumn1c{Source}&\\
\hline 
\endhead
\hline
\noalign{\vskip3pt}
\multicolumn{13}{r}{\itshape $T=4$ cont'd\dots}\\
\endfoot
\noalign{\vskip3pt}\hline
\endlastfoot
 & $^{64}$Ni       & -67098&.5&      0&.5&      &&&&            0+  &   4  &     07 &\\
 & $^{64}$Cu$^{i}$ & -58598&&        6 &&~~~~~6826&&       6&&  0+ & frg  & 07 & \footnote{strongest frg (cross section 100); other 16 keV lower (xs 37).}    
\end{longtable}

\newpage
\section{IMME coefficients}\label{IMMEtables}
\LTright=0pt
\LTleft=0pt
\tiny
\begin{longtable}{@{\extracolsep\fill}lcc}
\caption{$T=1$ IMME coefficients.}\label{tab:Teq1coeffs}\\\\
\noalign{\vskip-10pt}
\hline
  ~~~~~~~a (keV)     & b (keV)  & c (keV)   \\[3pt] 
  \hline\noalign{\vskip3pt}
\endhead
A = 6  \\   17649.7598 $\pm$  0.1000   &       -391.4551 $\pm$  2.5001   &      333.7852 $\pm$  2.5021    \\
A = 8  \\   21568.0000 $\pm$  3.0000   &       -987.8994 $\pm$  0.5006   &      365.7002 $\pm$  3.0415     \\   
A = 10 \\    13790.7998 $\pm$  0.4000   &      -1545.6548 $\pm$  0.2040   &      362.3452 $\pm$  0.4490     \\  
A = 12 \\    15108.0000 $\pm$  3.0000   &      -1984.3496 $\pm$  0.8201   &      245.7500 $\pm$  3.1101     \\
\hline
\end{longtable}


\clearpage
\LTright=0pt
\LTleft=0pt

\begin{longtable}{lccccc@{\extracolsep\fill}}
\caption{$T=3/2$ IMME coefficients.\\\label{tab:Teq3o2coeffs}}\\\\
\noalign{\vskip-10pt}
\hline
  ~~~~~~~a (keV) &  b (keV)  &  c (keV)  & d (keV)  &   $\chi^{2}$/dof  & prob \\[3pt] 
  \hline\noalign{\vskip3pt}
\endfirsthead
 \multicolumn{6}{@{}c@{}}{$T=3/2$ cont'd\dots \newline }\\
 \hline 
  ~~~~~~~a (keV) &  b (keV)  &  c (keV)  & d (keV)  &   $\chi^{2}$/dof  & prob \\[3pt] 
  \hline\noalign{\vskip3pt}
\endhead
 \hline
 \noalign{\vskip3pt}
 \multicolumn{6}{r}{ \itshape $T=3/2$ cont'd\dots}\\
\endfoot
 \noalign{\vskip3pt}\hline
\endlastfoot 
A = 7  \\ 
 26396.4681 $\pm$  23.9197   &    -537.3320 $\pm$   8.5693   &      214.1276 $\pm$  12.4257   &    -    &     2.275/1 & 0.132 \\ 
 26396.8750 $\pm$  23.9212   &    -608.1667 $\pm$  47.7422   &      212.5000 $\pm$  12.4725   &       32.6667 $\pm$  21.6596   & - &  
\\\\
A = 9  \\ 
 26337.1449 $\pm$   1.5952   &   -1319.4389 $\pm$   0.6833   &      265.2870 $\pm$   0.8519   &    -    &    20.194/1 & $< 0.001$ \\ 
 26339.0062 $\pm$   1.6481   &   -1333.8866 $\pm$   3.2871   &      263.9751 $\pm$   0.9006   &        6.7497 $\pm$   1.5021   & - &  
\\\\
\end{longtable}

\clearpage
\LTright=0pt
\LTleft=0pt
\tiny
\begin{longtable}{@{\extracolsep\fill}lcccccc}
 \caption{$T=2$ IMME coefficients.\\\label{tab:Teq2coeffs}}\\\\
\noalign{\vskip-10pt}
\hline
~~~~~~~a (keV) &  b (keV)  &  c (keV)  & d (keV)  &  e (keV) & $\chi^{2}$/dof  & prob \\[3pt]  
  \hline\noalign{\vskip3pt}
\endfirsthead
 \multicolumn{7}{@{}c@{}}{$T=2$ cont'd\dots}\\
 \hline 
 ~~~~~~~a (keV) &  b (keV)  &  c (keV)  & d (keV)  &  e (keV) & $\chi^{2}$/dof  & prob \\[3pt]  
  \hline\noalign{\vskip3pt}
\endhead
 \hline
 \noalign{\vskip3pt}
 \multicolumn{7}{r}{\itshape $T=2$ cont'd\dots}\\
\endfoot
 \noalign{\vskip3pt}\hline
\endlastfoot
A = 8  \\ 
 32433.6174 $\pm$ 1.8855 &  -875.1719 $\pm$ 3.2490 & 231.6009 $\pm$ 1.6984 &  -  & - &                  27.988/ 2  & $< 0.001$ \\ 
 32435.5728 $\pm$ 1.9258 &  -899.8053 $\pm$ 5.8955 & 224.5958 $\pm$ 2.2024 & 9.4168 $\pm$ 1.8797 & -    & 0.600/1 & 0.439 \\
 32436.0000 $\pm$ 1.0264 &  -875.6920 $\pm$ 0.5962 & 209.8026 $\pm$ 0.3027 &  -  & 5.3658 $\pm$ 0.0754 &  15.204/1 & $< 0.001$ \\ 
 32436.0000 $\pm$ 0.9684 &  -897.4734 $\pm$ 0.5594 & 219.5967 $\pm$ 0.2818 & 8.4733 $\pm$ 0.1404 & 1.4033 $\pm$  0.0703 &  - & -
\\\\
A = 12  \\ 
 27594.7594 $\pm$   2.3845   &   -1710.7285 $\pm$   5.6660   &      226.1032 $\pm$   2.9021   &    -     &    -    &    1.449/2 & 0.485 \\
\end{longtable}
\vskip20pt
\LTright=0pt
\LTleft=0pt
\tiny
\begin{longtable}{@{\extracolsep\fill}lcc}
 \caption{$T=5/2$ IMME coefficients.\\\label{tab:Teq5o2coeffs}}\\\\
\noalign{\vskip-10pt}
\hline
~~~~~~~a (keV) &  b (keV)  &  c (keV) \\[3pt] 
  \hline\noalign{\vskip3pt}
\endfirsthead
 \multicolumn{3}{@{}c@{}}{\bf $T=5/2$ cont'd\dots}\\
 \hline 
~~~~~~~a (keV) &  b (keV)  &  c (keV) \\[3pt]
  \hline\noalign{\vskip3pt}
\endhead
 \hline
 \noalign{\vskip3pt}
 \multicolumn{3}{r}{\itshape $T=5/2$  cont'd\dots}\\
\endfoot
 \noalign{\vskip3pt}\hline
\endlastfoot
A = 11  \\
 43172.8628 $\pm$   8.6748   &   -1592.7000 $\pm$   3.5745   &      245.9500 $\pm$   1.4030   
\\
A = 35  \\
-11290.6121 $\pm$  17.7099   &   -5860.3333 $\pm$  13.7437   &      173.3831 $\pm$   4.7379   
\\ 
A = 39  \\
-15108.6380 $\pm$  12.4113   &   -6272.3333 $\pm$   9.6206   &      158.2835 $\pm$   3.3271   
\\ 
A = 43  \\
-20480.7134 $\pm$  18.1195   &   -6898.6667 $\pm$   6.8398   &      184.3171 $\pm$   4.8326   
\\
A = 47  \\
-26793.2846 $\pm$   7.6495   &   -7464.7995 $\pm$   8.0044   &      179.1491 $\pm$   2.0787  
\\ 
A = 51  \\
-33261.0442 $\pm$  19.7704   &   -7929.3333 $\pm$  20.0689   &      140.9085 $\pm$   5.2751   
\end{longtable}

\vskip20pt
\LTright=0pt
\LTleft=0pt
\tiny
\begin{longtable}{@{\extracolsep\fill}lccccc}
 \caption{$T=3$ IMME coefficients.\\\label{tab:Teq3coeffs}}\\\\
\noalign{\vskip-10pt}
\hline
~~~~~~~a (keV) &  b (keV)  &  c (keV)  & d (keV)  & $\chi^{2}$/dof  & prob \\[3pt]  
  \hline\noalign{\vskip3pt}
\endfirsthead
 \multicolumn{6}{@{}c@{}}{\bf $T=3$ cont'd\dots}\\
 \hline 
~~~~~~~a (keV) &  b (keV)  &  c (keV)  & d (keV) & $\chi^{2}$/dof  & prob \\[3pt]  
  \hline\noalign{\vskip3pt}
\endhead
 \hline
 \noalign{\vskip3pt}
 \multicolumn{6}{r}{\itshape $T=3$ cont'd\dots}\\
\endfoot
 \noalign{\vskip3pt}\hline
\endlastfoot
A = 38  \\ 
 -9911.0000 $\pm$ 115.4686   &   -6218.5000 $\pm$ 122.2753   &      189.5000 $\pm$  28.5175   &     -    &    - & - 
\\\\
A = 46  \\ 
-22841.0684 $\pm$  11.5658   &   -7324.0834 $\pm$  11.1611   &      186.0721 $\pm$   2.6626   &     -    &    0.724/1 & 0.395 \\ 
-22833.8594 $\pm$  14.3473   &   -7324.8906 $\pm$  11.2194   &      181.3398 $\pm$   6.1687   &        1.4102 $\pm$   1.6575 &   -   & -  
\\\\ 
A = 50  \\ 
-29314.4706 $\pm$  17.0085   &   -7898.3985 $\pm$  14.2292   &      175.4409 $\pm$   2.8591   &     -    &    0.082/1 & 0.774 \\ 
-29311.0295 $\pm$   2.3108   &   -7906.4290 $\pm$   1.4911   &      180.3202 $\pm$   0.5349   &       -0.8616 $\pm$   0.1404 &   -   & -  
\end{longtable}
\end{appendix}

\end{document}